\documentclass[epj]{svjour}

\usepackage{graphicx}
\usepackage{fancyhdr}
\def\makeheadbox{{
 }}
\def\mytitle{My title} 
\def\myauthors{My name}  
\def\mytype{My type of session}
\def\mysession{My session}
\def\mytitle{Studies of penguin dominated $B$ decays at Belle} 
\def\myauthors{Yosuke Yusa}    
\def\mytype{Contributed Talk}    
\def\mysession{Flavor Physics}
\begin{document}
\title{Studies of penguin dominated $B$ decays at Belle}
\author{Y. Yusa\inst{1}
\thanks{\emph{Email:yyusa@vt.edu}}%
 \and
 Belle Collaboration
}
\institute{Virginia Polytechnic Inst. and State Univ.
}
\date{\today}
\abstract{
We present measurements of $CP$ violation parameter $\phi_{1}/\beta$ in $B^0$ decays that are dominated by $b \to c\bar{c}s$, $b \to sq\bar{q}$ and $b \to dq\bar{q}$ transitions. The results are based on a large sample of $B\bar{B}$ pairs recorded at the $\Upsilon(4S)$ resonance with the Belle detector at the KEKB energy-asymmetric $e^+e^-$ collider. $CP$ violation parameters for each decay mode are obtained from the asymmetries in the distributions of the proper-time intervals between the reconstructed $B$ and the accompanying $B$ meson.
\PACS{
      {PACS-key}{11.30.Er}   \and
      {PACS-key}{12.15.Hh}   \and
      {PACS-key}{13.25.Hw}
     } 
} 
\maketitle
%
\section{Introduction}
\label{intro}
The Standard Model (SM) describes $CP$ violation in $B^0$ meson decays using the complex phase of the $3 \times 3$ Cabbibo-Kobayashi-Masukawa (CKM) mixing matrix \cite{CKM}. In the decay chain $\Upsilon(4S) \to B^0 \bar{B}^0 \to f_{CP} f_{tag}$, where one of the $B$ mesons decays at time $t_{CP}$ to a final state $f_{CP}$ and the other decays at time $t_{tag}$ to a final state $f_{tag}$ that distinguishes between $B^0$ and $\bar{B}^0$, the decay rate has time dependence \cite{tCPV} given by 
$\displaystyle \mathcal{P}(\Delta t) = \frac{e^{-|\Delta t|/\tau_{B^0}}}{4\tau_{B^0}} \big[ 1+ q [ \mathcal{S}_f \sin(\Delta m_d \Delta t) + \mathcal{A}_f \cos(\Delta m_d \Delta t) ] \big]$.
\\
Here $\mathcal{S}_f$ and $\mathcal{A}_f$ are $CP$-violation parameters, $\tau_{B^0}$ is the $B^0$ lifetime, $\Delta m_d$ is the mass difference between the two $B^0$ mass eigenstates, $\Delta t = t_{CP} - t_{\rm tag}$, and the $b$-flavor charge is $q = +1(-1)$ when the tagging $B$ meson is a $B^0 (\bar{B}^0)$. 
To a good approximation, the SM predicts $\mathcal{S}_f = -\xi_f \sin 2 \phi_{1}$ and $\mathcal{A}_f = 0$ for both $b \to c\bar{c}s$ and $b \to sq\bar{q}$ transitions, where $\xi_f = +1(-1)$ corresponds to $CP$-even (-odd) final states. Recent theoretical studies within the SM framework \cite{phi1eff} find that the effective $\sin 2\phi_1$ values, $\sin 2\phi^{\rm eff}_1$, obtained from $b \to sq\bar{q}$ are expected to agree within $\mathcal{O}(0.01)$ with $\sin 2\phi_1$ from the $b \to c\bar{c}s$ transition. A comparison of $CP$-violation parameters between these theoretically clean $b \to sq\bar{q}$ modes and $b \to c\bar{c}s$ decays is an important test of the SM.
On the other hand, $\mathcal{S}_f$ is expected to be small for $b \to dq\bar{q}$ transition because the weak and quark mixing phases cancel \cite{S_dqq}.
In this paper, we report measurement of time-dependent $CP$-asymmetries of penguin dominated $B$ meson decays. Among the final states, $\phi K^0_S$, $\eta' K^0_S$, $\omega K^0_S$, $K^0_S \pi^0$ and $J/\psi K^0_S$ are $CP$ eigenstates with $\xi_f = -1$, while $\phi K^0_L$, $\eta' K^0_L$, $f_0 K^0_S$, $K^0_S \pi^0 \pi^0$, $K^0_S K^0_S K^0_S$ and $J/\psi K^0_L$ are $CP$ eigenstates with $\xi_f = +1$. Since $B^0 \to K^+ K^- K^0_S$ is a $CP$-even and -odd mixture, of both it has $\xi_f = -(2f_+ - 1)$, where $f_+$ is the $CP$-even fraction, measured to be $0.93 \pm 0.09 \pm 0.05$ assuming isospin relation \cite{CPfraction-KKKS}. The $CP$ asymmetry parameters for $b \to dq\bar{q}$ transition are measured from $B^0 \to K^0_S K^0_S$ decay mode.

\section{Measurement of $CP$ violation parameters}
\label{Meas}
At the KEKB energy-asymmetric $e^+e^-$ (3.5 on 8.0 GeV) collider \cite{KEKB}, the $\Upsilon (4S)$ is produced with a Lorentz boost of $\beta\gamma = 0.425$ nearly along the electron beam line, ($z$-axis). Since the $B^0$ meson pair is approximately at rest in the $\Upsilon (4S)$ center-of-mass system (cms), $\Delta t$ can be determined from the displacement in $z$ between the $f_{CP}$ and $f_{\rm tag}$ decay vertices:$\Delta t \simeq (z_{CP} - z_{\rm tag})/(\beta \gamma c) \equiv \Delta z/(\beta \gamma c)$.
The Belle detector \cite{belle} is a large-solid-angle magnetic spectrometer that consists of a silicon vertex detector (SVD), a 50-layer central drift chamber (CDC), an array of aerogel threshold Cherenkov counters (ACC), a barrel-like arrangement of time-of-flight scintillation counters (TOF), and an electromagnetic calorimeter comprised of CsI(Tl) crystals (ECL) located inside a super-conducting solenoid coil that provides a 1.5 T magnetic field. An iron flux-return located outside of the coil is instrumented to detect $K^0_L$ mesons and to identify muons (KLM). 

The intermediate meson states are reconstructed from the following decays: $\pi^0 \to \gamma\gamma$, $K_S^0 \to \pi^+\pi^-$ (denoted by $K_S^{+-}$ hereafter) or $\pi^0\pi^0$ (denoted by $K_S^{00}$ hereafter), $\eta \to \gamma\gamma$ or $\pi^+\pi^-\pi^0$, $\rho \to \pi^+\pi^-$, $\eta' \to \rho^0 \gamma$ or $\eta \pi^+\pi^-$, $\omega \to \pi^+\pi^- \pi^0$, $f_0 \to \pi^+ \pi^-$, $\phi \to K^+K^-$ and $J/\psi \to \ell^+\ell^- (\ell = \mu, e)$. We use all combinations of the intermediate states except for the following cases: $\eta \to \pi^+\pi^-\pi^0$ candidates are not used for $B^0 \to \eta' K^{00}_S$ decays; $\eta'\to\rho^0\gamma$ candidates are not used for $B^0 \to \eta' K^{0}_L$ decays; $K^{00}_S$ candidates are not used for $B^0 \to J/\psi K^0_S$, $f_0 K^0_S$, $\omega K^0_S$, $K^0_S \pi^0$, $K^0_S \pi^0 \pi^0$ and $K^0_S K^0_S$ decays. We reconstruct the $B^0\to K^0_S K^0_S K^0_S$ decays in the $K^{+-}_S K^{+-}_S K^{+-}_S$ or $K^{+-}_S K^{+-}_S K^{00}_S$ final states. In addition, $\phi \to K^{+-}_S K^{0}_L$ decays are used for the $B^0 \to \phi K_S^{+-}$ mode.

For reconstructed $B^0 \to f_{CP}$ candidates without a $K_L^0$ meson, $B$ meson decays are identified using the energy difference $\Delta E \equiv E^{\rm cms}_B - E^{\rm cms}_{\rm beam}$ and the beam-energy constrained mass $M_{\rm bc} \equiv \sqrt{(E^{\rm cms}_{\rm beam})^2 - (p^{\rm cms}_B)^2}$, where $E^{\rm cms}_B$ and $p^{\rm cms}_B$ are the cms energy and momentum of the reconstructed $B$ candidate, respectively. 
The signal candidates are selected by requiring 5.27 GeV/$c^2 < M_{\rm bc} <$ 5.29 GeV/$c^2$ and a mode-dependent $\Delta E$ window. Only $M_{\rm bc}$ is used to identify the decay $B^0 \to \phi K^0_S$ followed by $\phi \to K^0_S K^0_L$. Other candidate $B^0 \to f_{CP}$ decays with $K_L^0$ are selected using $p^{\rm cms}_B$, requiring 0.2 GeV/$c < p^{\rm cms}_B <$ 0.45 GeV/$c$ for $B ^0\to J/\psi K^0_L$ candidates and 0.2 GeV/$c < p^{\rm cms}_B <$ 0.5 GeV/$c$ for the others.

The dominant background for this analysis comes from continuum events: $e^+e^- \to u\bar{u}, d\bar{d}, s\bar{s}$, or $c\bar{c}$. To distinguish these topologically jet-like events from the spherical-like $B$ decay signal events, we combine a set of variables \cite{Belleccs05,SFW} that characterize the event topology into a signal (background) likelihood variable $\mathcal{L}_{\rm sig (bkg)}$, and impose loose mode-dependent requirements on the likelihood ratio $\mathcal{R}_{\rm s/b} \equiv \mathcal{L}_{\rm sig} /(\mathcal{L}_{\rm sig} + \mathcal{L}_{\rm bkg} )$. 

The contributions from $B\bar{B}$ events to the background for $B^0 \to f_{CP}$ candidates with a $K^0_L$ are estimated with Monte Carlo (MC) simulated events. The small $B\bar{B}$ contribution to the background in the $B^0 \to \eta' K^0_S$ mode is also estimated using MC events. We reject $K^0_S K^0_S K^0_S$ candidates if they are consistent with $B^0 \to \chi_{c0} K^0_S \to (K^0_S K^0_S) K^0_S$ or $B^0 \to D^0 K^0_S \to (K^0_S K^0_S) K^0_S$ decays, i.e., if one of the $K^0_S K^0_S$ pairs is consistent with the $\chi_{c0}$ mass or $D^0$ mass.
We estimate the contribution in the $B^0 \to \phi K^0_S$ sample from $B^0 \to K^+ K^- K^0_S$ and $B^0 \to f_0 K^0_S (f_0 \to K^+ K^-)$ decays from the Dalitz plot for $B^0 \to K^+ K^- K$ candidates with a method that is described elsewhere \cite{KKKDalitz}. The fraction of $B^0 \to K^+ K^- K^0_S$ events in the $B^0 \to \phi K^0$ sample is 2.75$\pm$0.14\%. The background from $B^0 \to f_0 K^0_S$ decay for $B^0 \to K^+ K^- K^0_S$ and $B^0 \to \phi K^0$ is found to be consistent with zero and the influence is treated as a source of systematic uncertainty.

The $b$-flavor of the accompanying $B$ meson is identified from inclusive properties of particles that are not associated with the reconstructed $B^0 \to f_{CP}$ decay. We use two parameters, the $b$-flavor charge $q$ and $r$, to represent the tagging information \cite{wrongTag}. The parameter $r$ is an event-by-event, MC-determined flavor-tagging dilution factor that ranges from $r = 0$ for no flavor discrimination to $r = 1$ for unambiguous flavor assignment. 

The vertex position for the $f_{CP}$ decay is reconstructed using charged tracks that have enough SVD hits \cite{vertex}. The $f_{\rm tag}$ vertex is obtained with well reconstructed track that are not assigned to $f_{CP}$. A constraint on the interaction-region profile in the plane perpendicular to the beam axis is also used with the selected tracks.

\begin{figure*}
\begin{center}
\includegraphics[width=0.27\textwidth,height=0.45\textwidth,angle=0]{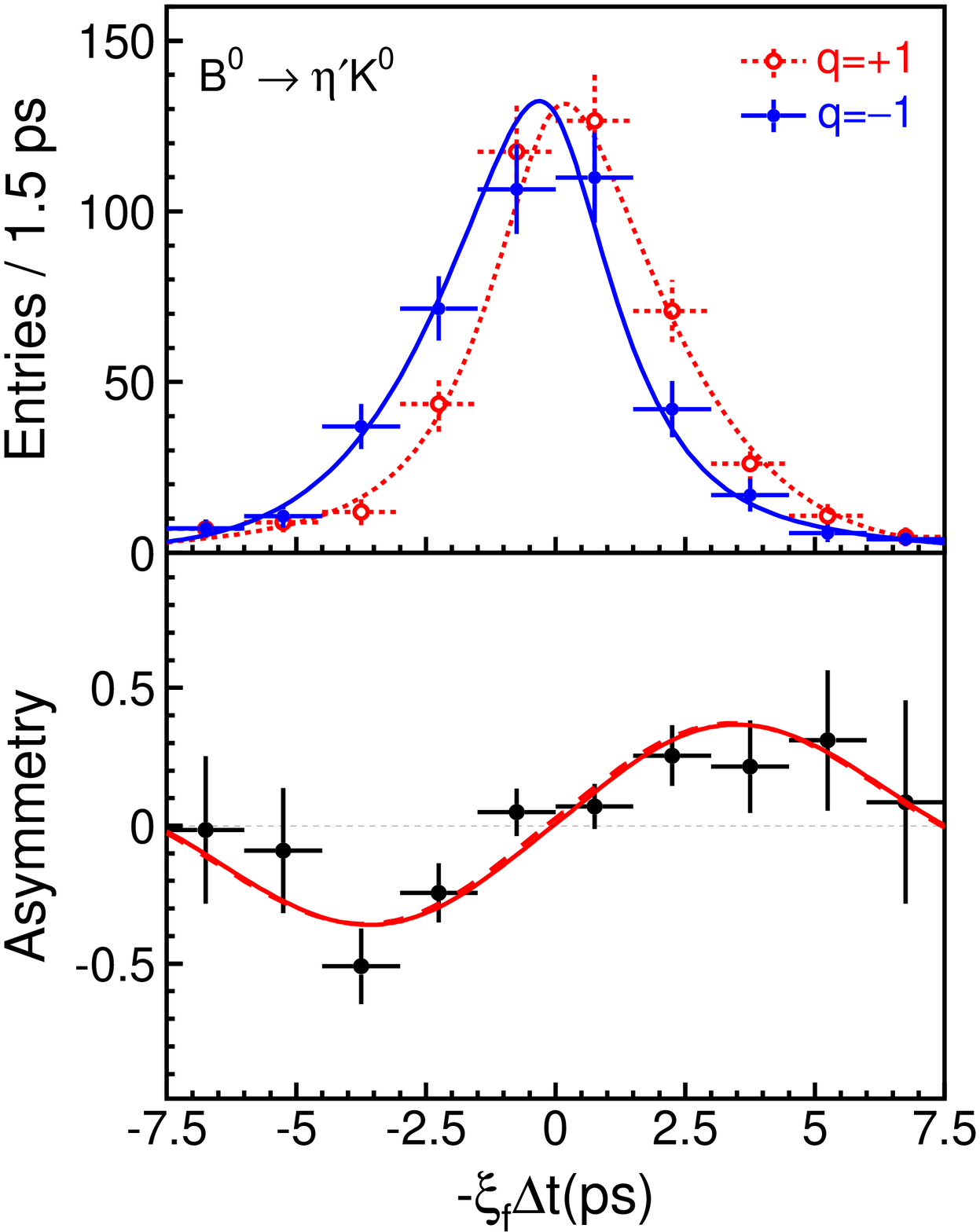}\includegraphics[width=0.27\textwidth,height=0.45\textwidth,angle=0]{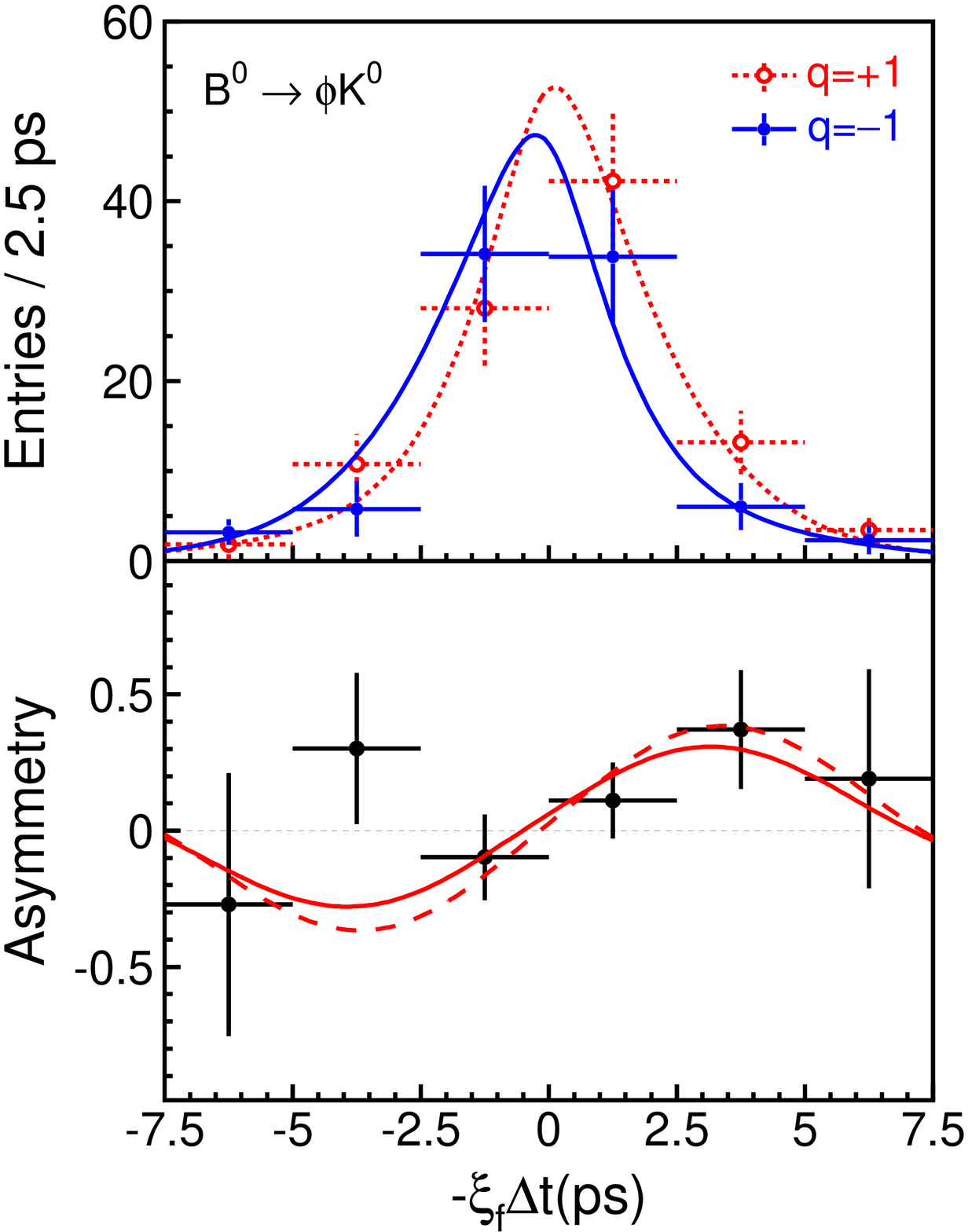}\includegraphics[width=0.27\textwidth,height=0.45\textwidth,angle=0]{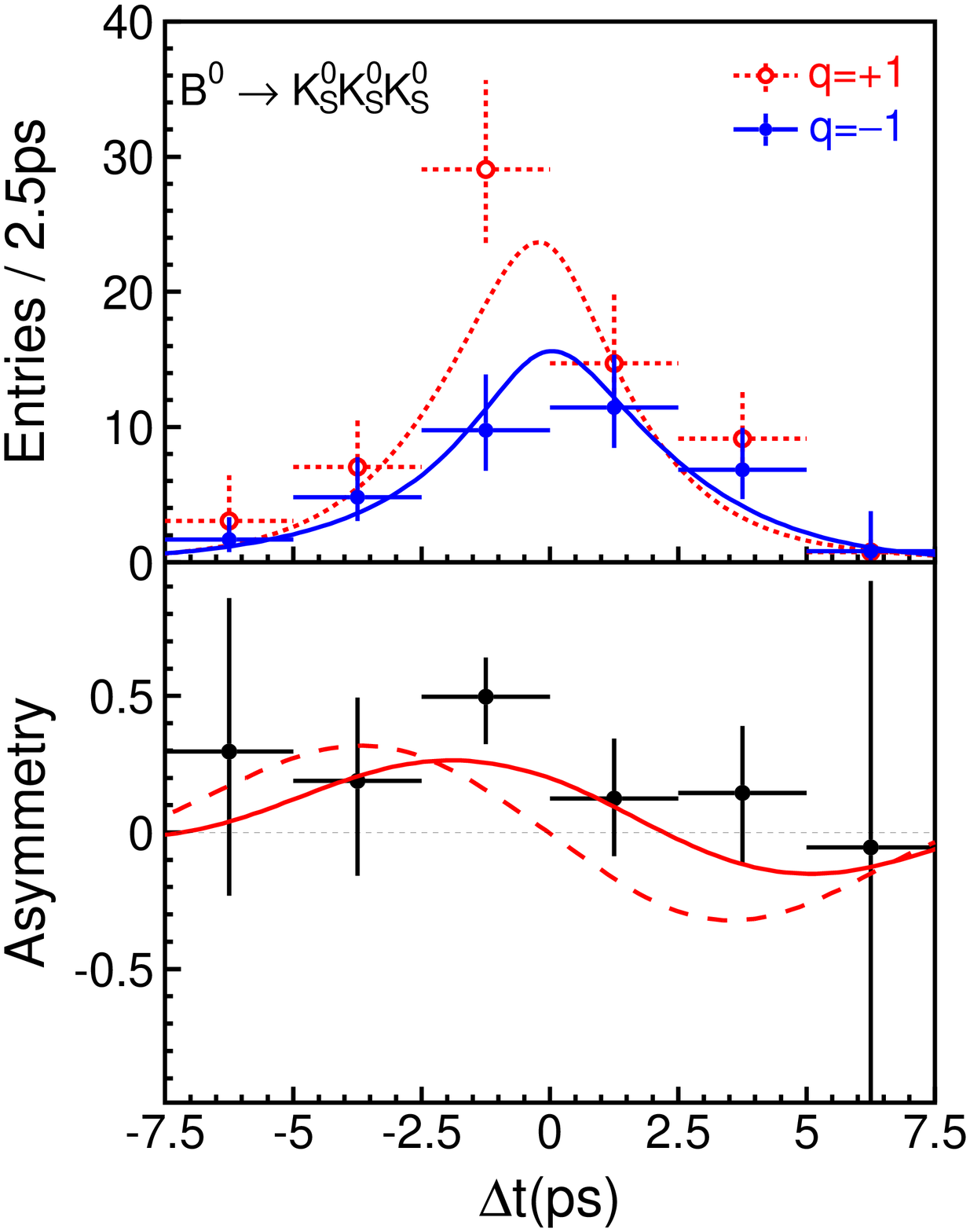}\includegraphics[width=0.27\textwidth,height=0.45\textwidth,angle=0]{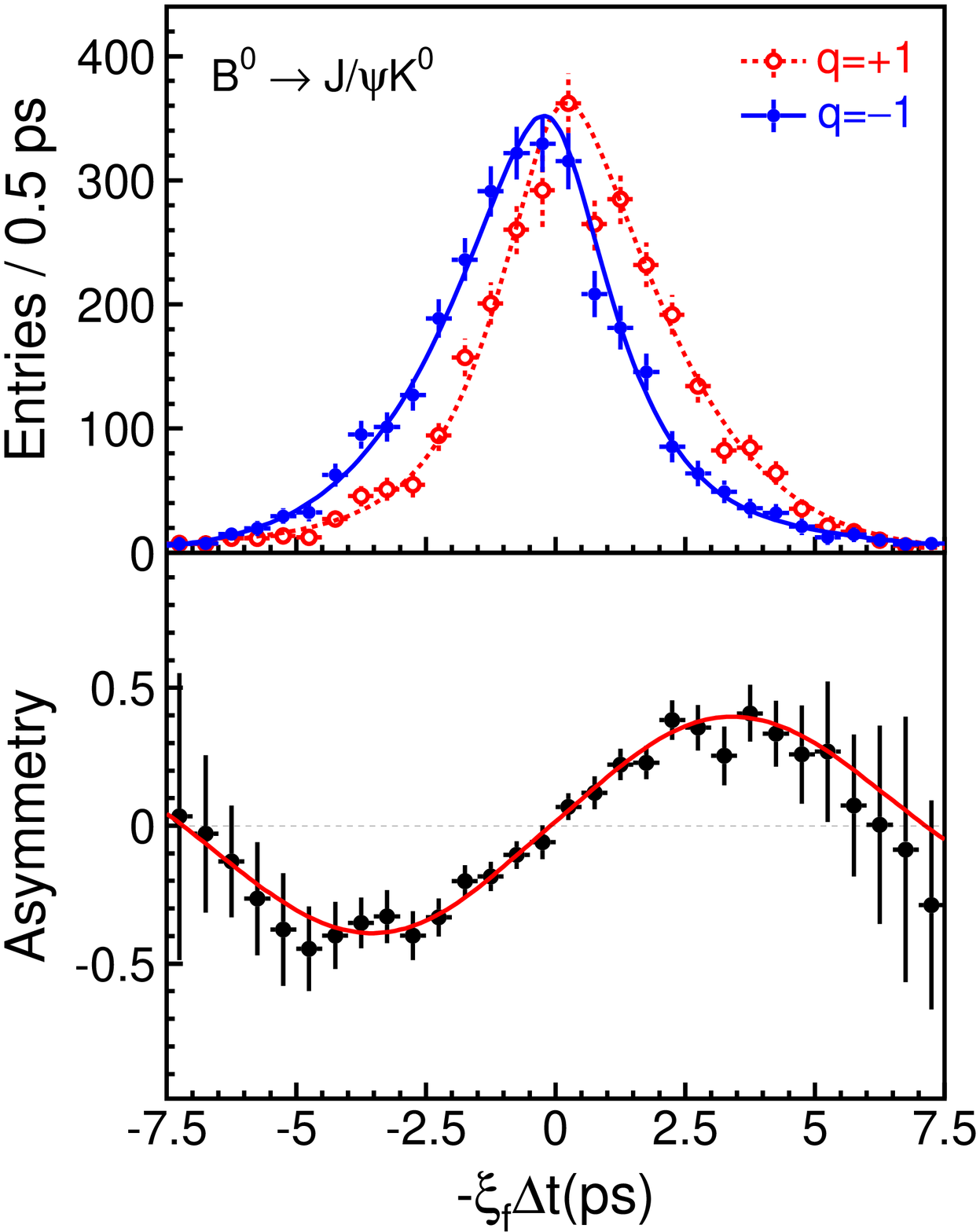}
\caption{Background subtracted $\Delta t$ distributions and asymmetries for event with good tag ($r > 0.5$) for $B^0 \to \eta' K^0$, $\phi K^0$, $K^0_S K^0_S K^0_S$ and $J/\psi K^0$. Dashed lines show the SM expectation from $B^0 \to J/\psi K^0$ measurement.}\label{fig_asym}
\end{center}
\end{figure*}

\begin{figure*}
\begin{center}
\includegraphics[width=0.25\textwidth,height=0.25\textwidth,angle=0]{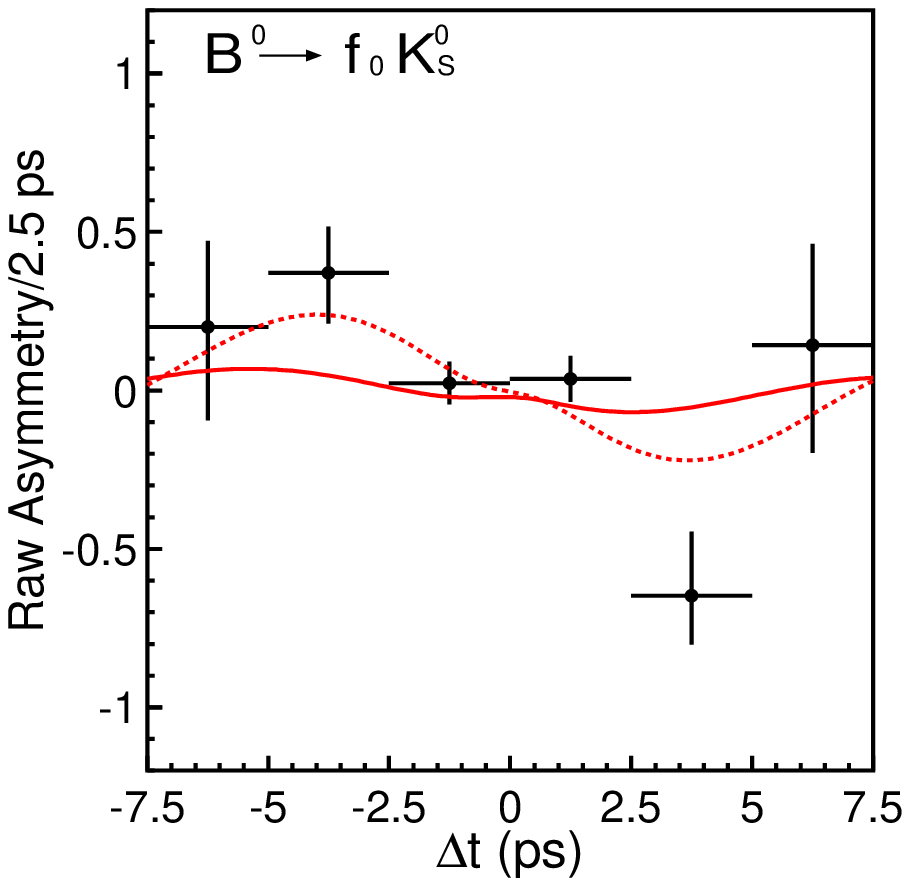}\includegraphics[width=0.25\textwidth,height=0.25\textwidth,angle=0]{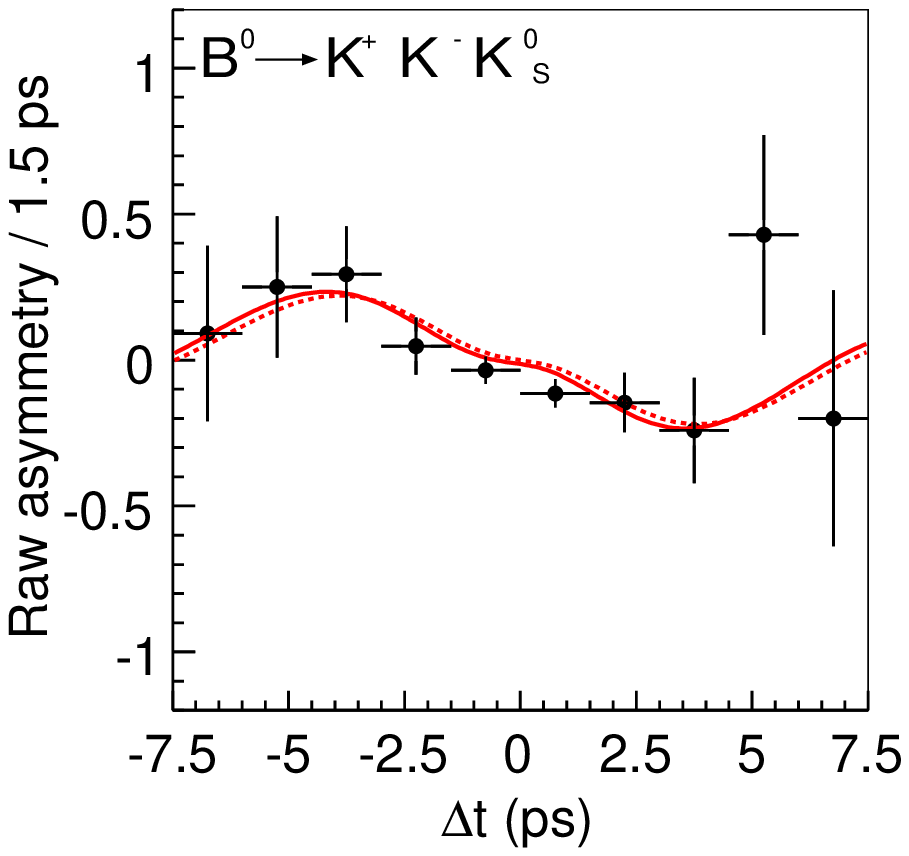}\includegraphics[width=0.25\textwidth,height=0.25\textwidth,angle=0]{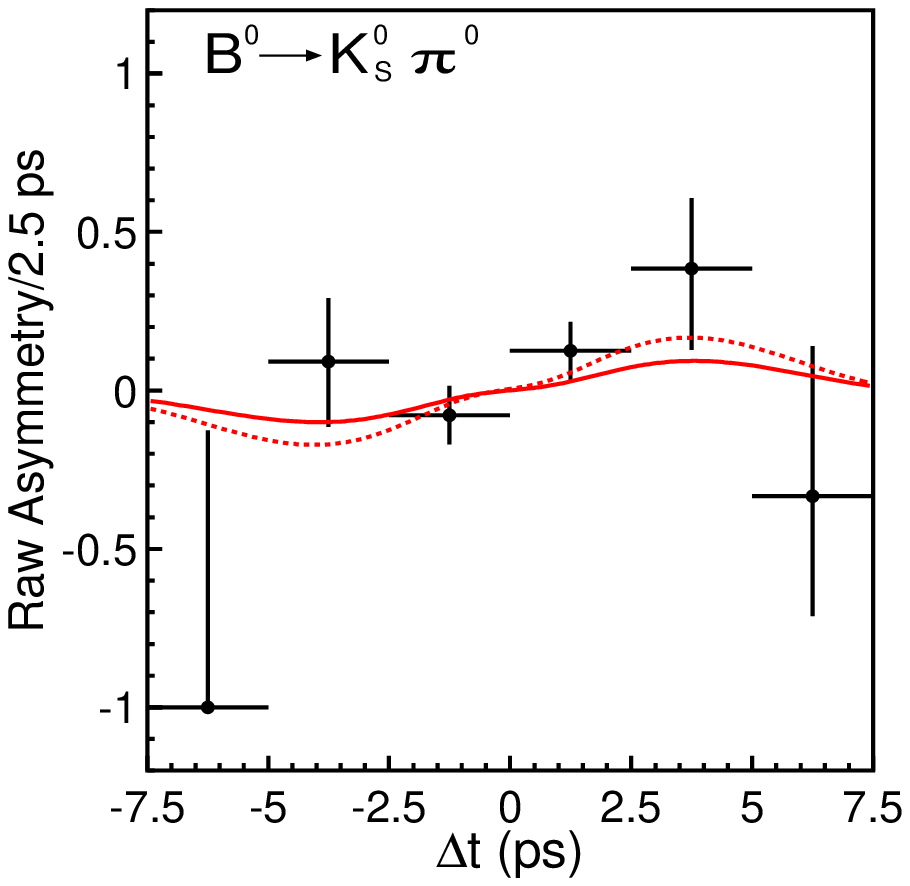}
\includegraphics[width=0.25\textwidth,height=0.25\textwidth,angle=0]{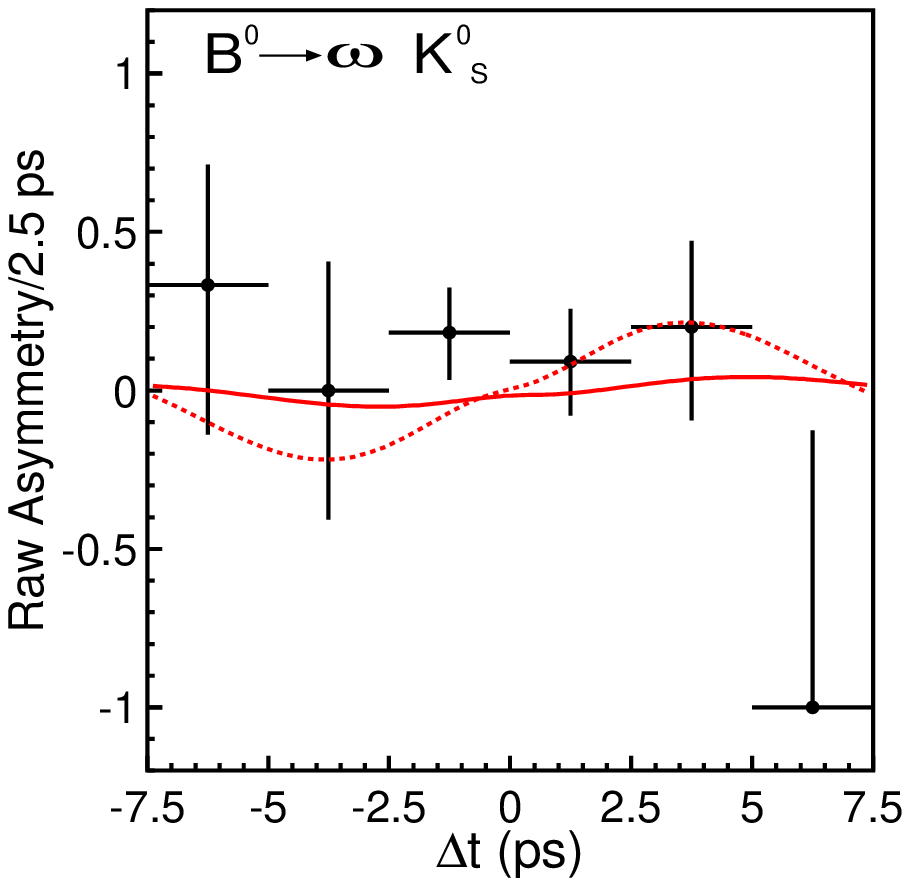}\includegraphics[width=0.26\textwidth,height=0.26\textwidth,angle=0]{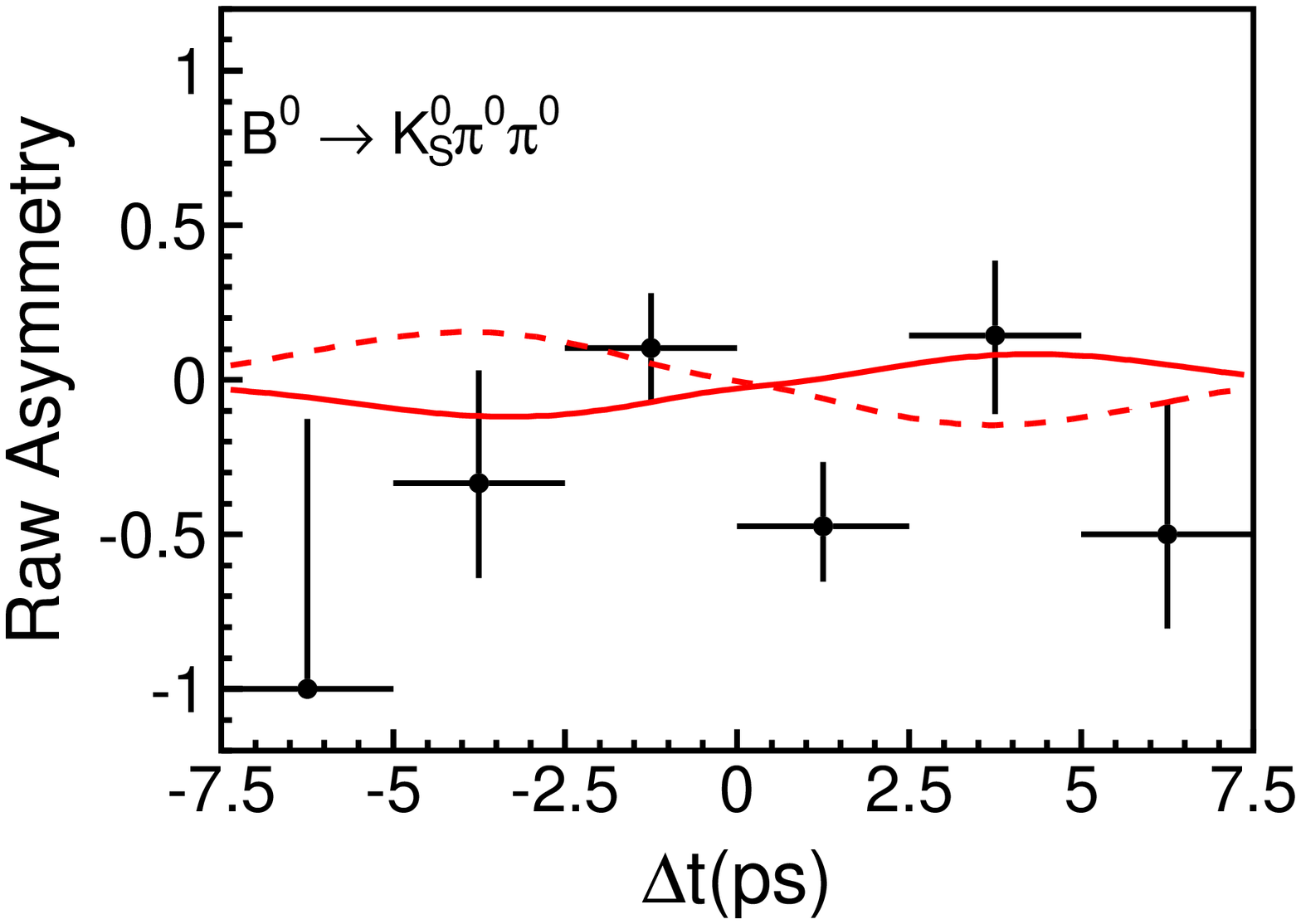}\includegraphics[width=0.26\textwidth,height=0.26\textwidth,angle=0]{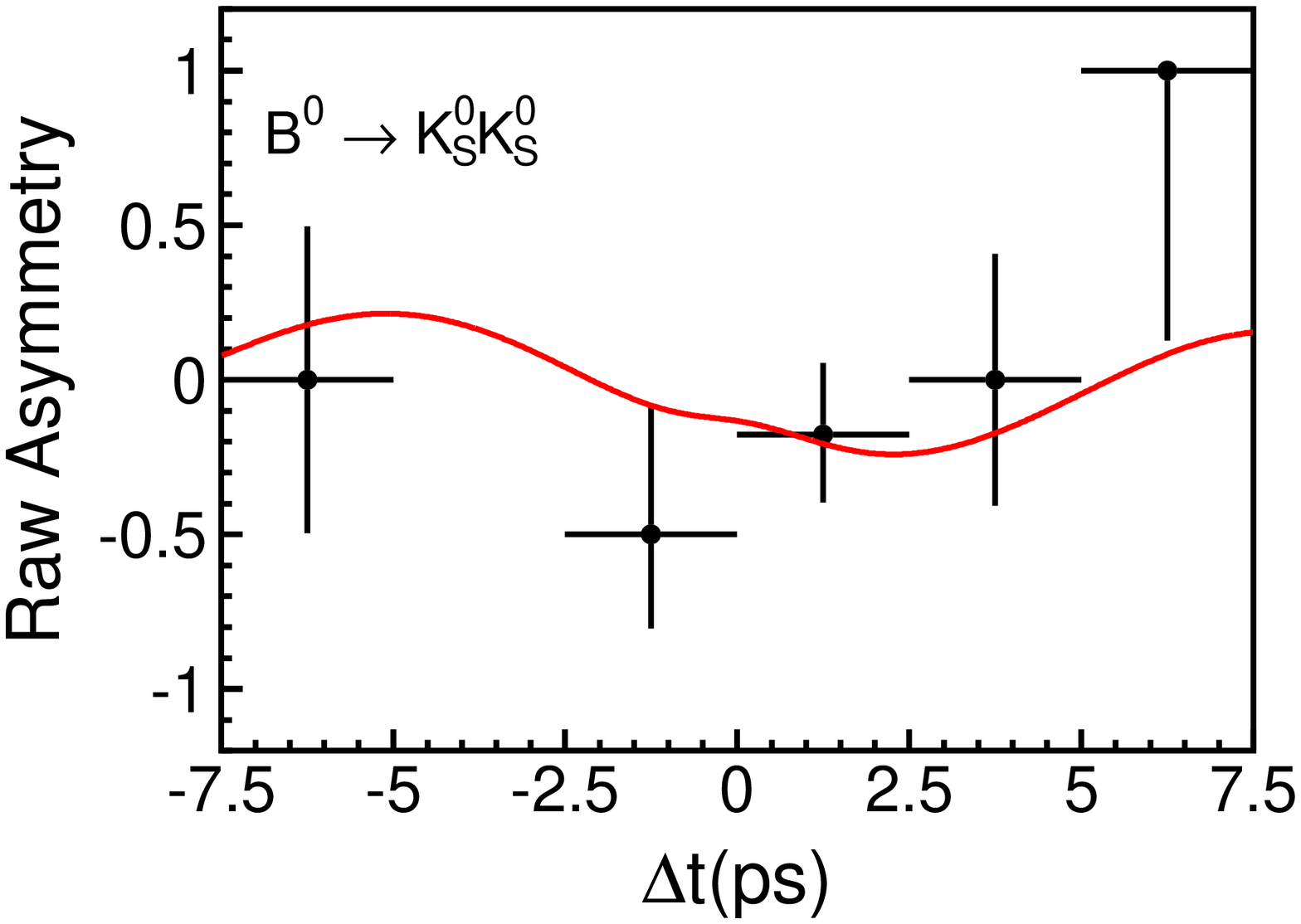}
\caption{Asymmetries for event with good tag ($r > 0.5$) for $B^0 \to f_0 K^0_S$, $K^+ K^- K^0$, $K^0_S \pi^0$, $\omega K^0_S$, $K^0_S \pi^0 \pi^0$ and $K^0_S K^0_S$. Dashed lines show the SM expectation from $B^0 \to J/\psi K^0$ measurement.}\label{fig_asym2}
\end{center}
\end{figure*}

We determine $CP$ parameters $\mathcal{S}_f$ and $\mathcal{A}_f$ for each mode by performing an unbinned maximum-likelihood fit to the observed $\Delta t$ distribution. The probability density function (PDF) expected for the signal distribution, $\mathcal{P}_{\rm sig}(\Delta t; \mathcal{S}_f , \mathcal{A}_f , q, w_l , \Delta w_l )$, is given by the time-dependent decay rate of signal 
incorporating the effect of incorrect flavor assignment. 
The distribution is convolved with the proper-time interval resolution function $\mathcal{R}_{\rm sig} (\Delta t)$, which takes into account the finite vertex resolution. The resolution and wrong-tag fractions are determined by a multi-parameter fit to the $\Delta t$ distribution of a high-statistics control sample of semileptonic and hadronic $b \to c$ decays \cite{Belleccs05,Bellesss05}. We determine the following likelihood for each event:
$P_i = (1 - f_{\rm ol}) \int \big[ f_{\rm sig} \mathcal{P}_{\rm sig} (\Delta t')R_{\rm sig} (\Delta t_i - \Delta t')
+ (1 - f_{\rm sig})\mathcal{P}_{\rm bkg}(\Delta t') R_{\rm bkg} (\Delta t_i - \Delta t')\big] d(\Delta t') + f_{\rm ol}P_{\rm ol}(\Delta t_i).$ 
The signal probability $f_{\rm sig}$ depends on the $r$ region and is calculated on an event-by-event basis as a function of the following variables: $\Delta E$ and $M_{\rm bc}$ for $B^0 \to J/\psi K^0_S$; $p^{\rm cms}_B$ for $B^0 \to J/\psi K^0_L$ and $\phi K^0_L$; $M_{\rm bc}$ and $\mathcal{R}_{\rm s/b}$ for $B^0 \to \phi (\to K^0_S K^0_L)$ $K^0_S$; $\Delta E$, $M_{\rm bc}$ and $\mathcal{R}_{\rm s/b}$ for the other modes. $\mathcal{P}_{\rm bkg}(\Delta t)$ is a PDF for background events, which is convolved with the background resolution function $R_{\rm bkg}$. The term $P_{\rm ol}(\Delta t)$ is a broad Gaussian function that represents a small outlier component \cite{Belleccs05,Bellesss05}. The 
$\mathcal{S}_f$ and $\mathcal{A}_f$ are determined by maximizing the likelihood function $L = \prod_i P_i(\Delta t_i; \mathcal{S}_f, \mathcal{A}_f)$ where the product is over all events. Table \ref{tab1} summarizes the fit results for $\sin 2 \phi^{\rm eff}_1$ and $\mathcal{A}_f$. 
Figures \ref{fig_asym} and \ref{fig_asym2} show the $\Delta t$ distributions and asymmetries for good tag quality events of $r>0.5$. 
The dominant sources of systematic error for $\sin 2\phi^{\rm eff}_1$ stem from the uncertainties in the resolution function and in the background fraction. The dominant sources for $\mathcal{A}_f$ are the effects of tag-side interference (TSI)\cite{tag-interference}, the uncertainties in the background fraction, in the vertex reconstruction and in the resolution function. We study the possible correlations between $\mathcal{R}_{\rm s/b}$, $p^{\rm cms}_B$ and $r$ PDFs used for $\phi K^0_L$ and $\eta' K^0_L$, which are neglected in the nominal result, and include their effect in the systematic uncertainties in the background fraction. Other contributions come from uncertainties in wrong tag fractions, the background $\Delta t$ distribution, $\tau_{B^0}$ and $\Delta m_d$. A possible fit bias is examined by fitting a large number of MC events and is found to be small. %
The dominant sources of systematic errors for the $B^0 \to J/\psi K^0$ mode are the uncertainties in the vertex reconstruction, in the resolution function, in the background fraction, in the flavor tagging, a possible fit bias, and the effect of the TSI. Other contributions are negligible. We add each contribution in quadrature to obtain the total systematic uncertainty. 
The systematic errors are summarized in Table \ref{tab2}.

\begin{table*}
\caption{Number of signal $N_{\rm signal}$ and results of the fits to $\Delta t$ distributions. The first errors are statistical and the second errors are systematic. The third error for $\sin 2 \phi^{\rm eff}_1$ of $K^+ K^- K^0_S$ mode is an additional systematic error arising from the uncertainty of the $\xi_f = +1$ fraction.}\label{tab1} 
\begin{tabular}{lccc}
\hline\noalign{\smallskip}
Mode                & $N_{\rm signal}$                              & $\sin 2 \phi^{\rm eff}_1$ & $\mathcal{A}_f$\\
\noalign{\smallskip}\hline\noalign{\smallskip}
$\phi K^0$          &  307$\pm$21 ($K^0_S$), 114$\pm$17 ($K^0_L$)   & $+0.50 \pm 0.21 \pm 0.06$ & $+0.07 \pm 0.15 \pm 0.05$\\
$\eta' K^0$         & 1421$\pm$46 ($K^0_S$), 454$\pm$39 ($K^0_L$)   & $+0.64 \pm 0.10 \pm 0.04$ & $-0.01 \pm 0.07 \pm 0.05$\\
$\omega K^0_S$      &  118$\pm$17                                   & $+0.11 \pm 0.46 \pm 0.06$ & $-0.09 \pm 0.29 \pm 0.06$\\
$K^0_S \pi^0$       &  515$^{+32}_{-31}$                            & $+0.33 \pm 0.35 \pm 0.08$ & $-0.05 \pm 0.14 \pm 0.05$\\
$K^0_S \pi^0 \pi^0$ &  307$\pm$32                                   & $-0.43 \pm 0.49 \pm 0.09$ & $-0.17 \pm 0.24 \pm 0.05$\\
$f_0 K^0_S$         &  377$\pm$25                                   & $+0.18 \pm 0.23 \pm 0.11$ & $-0.15 \pm 0.15 \pm 0.07$\\
$K^0_S K^0_S K^0_S$ &  185$\pm$17                                   & $+0.30 \pm 0.32 \pm 0.08$ & $+0.31 \pm 0.20 \pm 0.07$\\
$K^+ K^- K^0$       &  840$\pm$34                                   & $+0.68 \pm 0.15 \pm 0.03^{+0.23}_{-0.13}$    & $-0.09 \pm 0.10 \pm 0.05$\\
$K^0_S K^0_S$       &   58$\pm$11                                   & $\mathcal{S}_f = -0.38 \pm 0.77 \pm 0.08$    & $-0.38 \pm 0.38 \pm 0.05$\\
$J/\psi K^0$        & 7484$\pm$87 ($K^0_S$), 6512$\pm$123 ($K^0_L$) & $\sin 2 \phi_1 = +0.642 \pm 0.031 \pm 0.017$ & $+0.018 \pm 0.021 \pm 0.014$\\
\noalign{\smallskip}\hline
\end{tabular}
\vspace*{1cm}  
\end{table*}

\begin{table*}
\caption{Dominant source of systematic error.\label{tab2}} 
\begin{tabular}{lcccccccccc}
\hline\noalign{\smallskip}
                    & $\phi K^0$ & $\eta' K^0$  & $\omega K^0_S$ & $K^0_S \pi^0$ & $K^0_S \pi^0\pi^0$ & $f^0 K^0_S$ & $K^0_S K^0_S K^0_S$  & $K^+ K^- K^0$ & $K_S^0 K_S^0$ & $J/\psi K^0$\\
\noalign{\smallskip}\hline\noalign{\smallskip}
\multicolumn{6}{l}{$\sin 2\phi^{\rm eff}_1$}\\
Resolution function & 0.04       & 0.04         & 0.05           & 0.07          & 0.04               & 0.02        & 0.05                 & 0.08          & 0.06          & 0.006 \\
Background fraction & 0.04       & 0.02         & 0.04           & $<0.01$       & 0.05               & 0.04        & 0.06                 & 0.01          & 0.04          & 0.006\\
Background $\Delta t$ shape & 0.01 & $<0.01$    & $<0.01$        & 0.05          & 0.05               & 0.09        & 0.01                 & $<0.01$       & 0.04          & 0.001\\
\noalign{\smallskip}\hline\noalign{\smallskip}
\multicolumn{6}{l}{$\mathcal{A}_f$}\\
Resolution function   & 0.02     & 0.01         & 0.02           & $<0.01$       & 0.02               & $<0.01$     & 0.02                 & 0.05          & $<0.01$       & 0.001\\
Background fraction   & 0.04     & 0.02         & 0.02           & $<0.01$       & 0.03               & 0.03        & 0.06                 & 0.07          & 0.02          & 0.009\\
Vertex reconstruction & 0.02     & 0.02         & 0.02           & 0.02          & 0.02               & 0.02        & 0.02                 & 0.02          & 0.02          & 0.009\\
TSI \cite{tag-interference} & 0.03 & 0.02 & 0.04 & 0.04        & 0.04               & 0.04        & 0.04                 & 0.03          & 0.03          & 0.009 \\
\noalign{\smallskip}\hline
\end{tabular}
\vspace*{1cm}  
\end{table*}

\section{Summary}
\label{summary}
For the $B^0 \to \eta' K^0$ mode, we determine the statistical significance from the obtained statistical uncertainties, taking into account the effect of the systematic uncertainties. The Feldman-Cousins frequentist approach \cite{FeldmanCousins} gives the significance of $CP$ violation that is equivalent to 5.6 standard deviations for a Gaussian error.We conclude that we have observed mixing-induced $CP$ violation in the mode $B^0 \to \eta' K^0$. 
We do not find any significant difference between the results for each individual $b \to sq\bar{q}$ and $b \to dq\bar{q}$ mode and those predicted from SM. Since some models of new physics predict such effects, our results can be used to constrain these models. However, many models predict smaller deviations that we cannot rule out with the current experimental sensitivity. Therefore, further measurements with much larger data samples are required in order to search for new, beyond the SM, $CP$-violating phases in the $b \to s$ transition. 

We thank the KEKB group for excellent operation of the accelerator, the KEK cryogenics group for efficient solenoid operations, and the KEK computer group and the NII for valuable computing and Super-SINET network support. We acknowledge support from MEXT and JSPS (Japan); ARC and DEST (Australia); NSFC and KIP of CAS (China); DST (India); MOEHRD, KOSEF and KRF (Korea); KBN (Poland); MIST (Russia); ARRS (Slovenia); SNSF (Switzerland); NSC and MOE (Taiwan); and DOE (USA).

\end{document}